\documentclass{article}

\usepackage{arxiv}

\usepackage[utf8]{inputenc} 
\usepackage[T1]{fontenc}    
\usepackage{hyperref}       
\usepackage{url}            
\usepackage{booktabs}       
\usepackage{amsfonts}       
\usepackage{nicefrac}       
\usepackage{microtype}      
\usepackage{lipsum}
\usepackage{graphicx}
\usepackage{amsmath}
\usepackage{algorithm}
\usepackage{algorithmic}

\newcommand{\country}{Russia}

\newcommand{\mau}{100M}
\newcommand{\egovk}{Ego-VK}
\newcommand{\VK}{VK}
\newcommand{\aboutlink}{\url{https://vk.com/about}}
\newcommand{\mediumlink}{\url{https://vkteam.medium.com/pymk-at-vk-ml-over-ego-nets-b31f5df7e944}}
\newcommand{\gitlink}{\url{https://github.com/ezamyatin/walk_gnn}}
\newcommand{\nodetoveclink}{\url{https://github.com/ezamyatin/node2vec-spark}}

\title{GNN Applied to Ego-nets for Friend Suggestions}

\author{
 Evgeny Zamyatin \\
  VK\\
  Russia, Saint-Petersburg \\
  \texttt{e.zamyatin@corp.vk.com}
}

\begin{document}
\maketitle
\begin{abstract}
A major problem of making friend suggestions in social networks is the large size of social graphs, which can have hundreds of millions of people and tens of billions of connections. Classic methods based on heuristics or factorizations are often used to address the difficulties of scaling more complex models. However, the unsupervised nature of these methods can lead to suboptimal results. In this work, we introduce the Generalized Ego-network Friendship Score framework, which makes it possible to use complex supervised models without sacrificing scalability. The main principle of the framework is to reduce the problem of link prediction on a full graph to a series of low-scale tasks on ego-nets with subsequent aggregation of their results. Here, the underlying model takes an ego-net as input and produces a pairwise relevance matrix for its nodes. In addition, we develop the WalkGNN model which is capable of working effectively in the social network domain, where these graph-level link prediction tasks are heterogeneous, dynamic and featureless. To measure the accuracy of this model, we introduce the \egovk{} dataset that serves as an exact representation of the real-world problem that we are addressing. Offline experiments on the dataset show that our model outperforms all baseline methods, and a live A/B test demonstrates the growth of business metrics as a result of utilizing our approach.
\end{abstract}


\section{Introduction}
\label{introduction}

The link prediction task is one of the fundamental problems of graph analysis. It is widespread in the industry, especially in social networks. Many of their long-term metrics depend on the users' social activity, and recommender systems help support and promote this activity. The social graph of a modern service represents a heterogeneous continuous-time dynamic graph. Various connections between users appear and disappear, including friendships, messages, likes, profile visits and others. Many of them are characterized by timestamps and intensities. The task of building methods that can effectively take into account all the available information about user relationships is very important, as the solutions' quality can directly affect business performance.

One of the specifics of working with social networks is the lack of meaningful features that describe users. General attributes such as sex, age or interests are often available, but they are not sufficient to model a user's social dynamics. Some bioinformatics problems can be seen as examples of tasks with rich node features, where units have natural descriptions that precisely define the principles of link formation. In the case of social networks, the structural information of the graph is of particular importance.

Despite active research~\cite{Keramatfar22,Abadal20} on graph neural networks, heuristics-based approaches still play a significant role in the industry. Methods such as common neighbors, Adamic-Adar~\cite{AdamicA03}, Preferential Attachment~\cite{Barabasi99}, etc. are widely used. Many recent papers refer to them as strong baselines~\cite{ZhangC18,EpastoLMSTV15,PerozziAS14}, and the field of research aimed at finding ways to generalize them shows high potential~\cite{ZhangC18,ZhuZXT21,EpastoLMSTV15}. Most of the heuristics do not rely on node features, which is an important property for the task of link prediction in social networks. The scalability, simplicity and efficiency of such methods make them attractive for industry use.

Most popular heuristics rely on local neighborhood analysis, which makes it possible to scale them to large graphs. To compute them, storing the whole graph in memory is not needed, since some local structures are enough. For example, a distributed version of the triangle counting algorithm~\cite{SuriV11,Schank07}, which requires storing only the list of the node's neighbors, can be used to compute common neighbors and Adamic-Adar heuristics.

\begin{figure}[!h]
  \centering
  \includegraphics[width=0.5\linewidth]{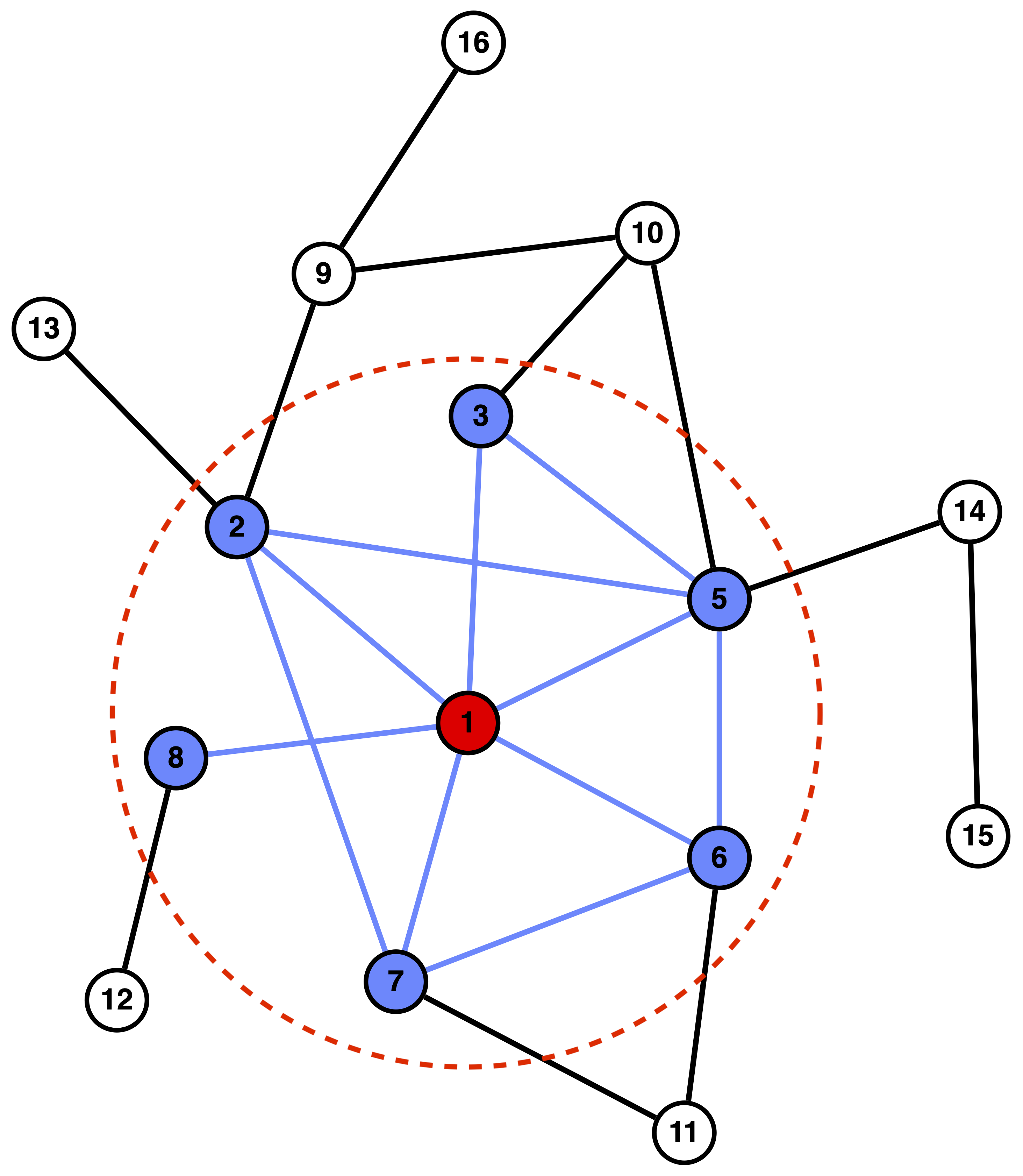}
  \caption{Ego-net example.}
  \label{fig:ego_net}
\end{figure}

One of the most informative and convenient structures for analyzing local neighborhoods is the ego-net. The ego-net of a node is a subgraph containing all the neighbors of the node and the links between them (see~\autoref{fig:ego_net}). On the one hand, ego-nets provide a rich description of local neighborhoods, and their small size makes it possible to use complex methods for their analysis. On the other hand, ego-nets can still be processed within large graphs, as there is an efficient way~\cite{EpastoLMSTV15,SuriV11} to construct them using MapReduce~\cite{DeanG08}. The maximum volume of all ego-nets within a graph is bound by $O(|E|^{3/2})$~\cite{Schank07}, which is acceptable for industrial graphs. All of the above makes ego-nets a potentially attractive structure for working with graphs that contain billions of nodes.

In this work, we propose a general framework for building recommendations on large-scale graphs, inspired by the idea on making friend suggestions using ego-net analysis~\cite{EpastoLMSTV15}. The main principle of the framework is to reduce the problem of link prediction in a whole graph to a series of tasks within ego-nets and the aggregation of their results. The task within an ego-net can be characterized as a graph-level link prediction problem in heterogeneous dynamic graphs without node features. This is yet another challenge for GNNs when it comes to inductively determining the laws of edge dynamics in graphs. The lack of meaningful node attributes in the social network domain make the graph structure the only available source of information. The graph structure, on the other hand, is highly descriptive, as there are many possible connections between users in social networks. 

The combination of these properties makes our task challenging and novel. First-order GNNs fail on this task because they strongly depend on the quality attributes of the nodes~\cite{ErricaPBM20}. Higher-order GNNs are theoretically capable of solving the problem, but their application to heterogeneous graphs with rich attributes of edges is not sufficiently covered in the literature. To address this problem, we introduce the WalkGNN model -- a second-order GNN, which constructs representations for pairs of nodes and is able to efficiently account for different types of links between nodes together with their numerical characteristics. Its key component is the WalkConv layer, which transforms each edge into an information filter and passes through it the corresponding state the relationship of a pair of nodes.

Our research is based on data from \VK{}\footnote{\aboutlink{}}, the largest social network in \country{} with \mau{} monthly active users. To measure the models' quality on our task, we introduce the \egovk{} dataset, which represents the set of ego-nets of \VK{} users. Each ego-net has at most $300$ nodes and $4$ possible types of connections between them. One of the edge types is the friendship's age, and the three other edges describe other types of relationships. The task is to predict new friendships in each ego-net. 

The rest of the paper is organized as follows. In Section 2,
we briefly review the literature relevant to our work. Then,
in Section 3, we present our Generalized Ego-network Friendship Score framework and describe the architecture of the WalkGNN model. In Section 4, we describe the experimental setting, the offline and online results, and the ablation study. Finally, in Section 5, we draw our conclusions and consider the directions for future research.

\section{Related works}

Traditionally, the problem of link prediction in large-scale graphs is solved by one of two types of methods. One is based on local neighborhood analysis using heuristics, while the other is based on graph embeddings construction. Approaches that use neural networks, in most cases, turn out to be too complex to scale and are not directly applicable to large-scale graphs. However, given that in our work we reduce the large-scale problem to a series of low-scale tasks, the work related to building complex models for the link prediction problem is also relevant here.

\textbf{Heuristic Methods.}
Despite the advanced age of most graph heuristics, they are still applicable due to their efficiency, simplicity and scalability. In contrast to most complex models, these methods do not require the presence of node features. The most popular among them are first- and second-order heuristics. Higher-order heuristics are much more difficult to scale on large graphs, so they much less commonly used in the industry. Most used first-order heuristics are Common Neighbours (CN) and Preferential Attachment (PA)~\cite{Barabasi99}. Adamic-Adar (AA)~\cite{AdamicA03} and resource allocation (RA)~\cite{Tao09} are examples of second-order heuristics. Murata et al.~\cite{MurataM07} and Lu et al.~\cite{lu2010link} also consider the weighted versions of these algorithms. All of these heuristics were invented a long time ago, but remain strong baselines for many link prediction problems.

Zhang et al.~\cite{ZhangC18} proposed the SEAL method, which can automatically learn heuristics via analyzing local subgraphs around each potential link. While this approach performed well, the relative uplift compared to the well-known Adamic-Adar was below 5\% in most cases. On the other hand, the disadvantage of SEAL is the difficulty of scaling it to large graphs.

Epasto et al.~\cite{EpastoLMSTV15} proposed an original idea of analyzing local communities within ego-nets to improve the common neighbors heuristic. Their friendship score similarity measure the co-occurrences of two nodes in different ego-net's clusters. This approach shows high accuracy and scales well to industrial graphs, but still relies on heuristic approaches. Since a variation of this method is deployed on \VK{}, we use it as a baseline in our online tests.

Taking the idea of link prediction using ego network analysis as a starting point, our work builds on it by exploring the use of supervised methods to obtain more accurate solutions.

\textbf{Embedding Methods.}
Most works on link prediction using large graphs belongs to this family of methods. Embedding methods learn a representation for each node by capturing the edge structure of the graph. A dot or cosine similarity measure is most commonly used for recommendations. The classical methods such as DeepWalk~\cite{PerozziAS14}, node2vec~\cite{GroverL16} and LINE~\cite{TangQWZYM15} have proven to be robust and efficient methods for link prediction in static homogeneous graphs. Meanwhile, in case of heterogeneous or dynamic graphs, these methods lose their effectiveness due to the unsupervised nature of learning. CTDNE~\cite{NguyenLRAKK18} extends the DeepWalk method to the case of continuous-time dynamic graphs and shows good results. However, the principle of temporal walks can essentially be considered heuristics rather than supervised learning, so it can be suboptimal. Various methods~\cite{BordesUGWY13,SunDNT19,YangYHGD14a,Kazemi018} for multi-relational knowledge graphs can be applied to heterogeneous graphs, but they are difficult to adapt for a temporal setting.

Embedding-based methods are attractive in terms of scalability,
but still not flexible enough for supervised learning. Our work
is not directly compared to these methods, but in this subsection
we show the importance of research on the application of supervised
approaches to large-scale graphs.

\textbf{Graph Neural Nets.}
Generally, complex transductive models do not scale well to large graphs due to the graphs' huge number of parameters and nonlinear structure. In this paper, we reduce the problem of link prediction in a complete graph to the low-scale graph-level problems on ego-nets, which is why we can use GNNs.

At present, the most popular way of constructing graph neural networks is the message-passing framework~\cite{GilmerSRVD17}. Its main idea is to hierarchically build representations of nodes through convolutions of neighbors. The GraphSAGE~\cite{HamiltonYL17}, TGAT~\cite{XuRKKA20}, GAT~\cite{VelickovicCCRLB18} and GCN~\cite{KipfW17} models are all different variations of MP-GNNs. They show good results on a number of problems, but fail at tasks with no natural node attributes. 

An active line of research is focused on evaluating the impact of node availability and quality of node characteristics on the accuracy of graph neural networks. Xu et al.~\cite{XuHLJ19}, Morris et al.~\cite{Morris19} and Dehmamy et al.~\cite{DehmamyBY19} consider the expressive power of message-passing GNNs~\cite{GilmerSRVD17} from a theoretical point of view. They prove that these types of models have significant limitations on learning from graph structures. Errica et al.~\cite{ErricaPBM20} and Duong et al.~\cite{Duong19} empirically demonstrate that the success of many GNNs relies heavily on rich node features. For example, in bioinformatics problems~\cite{HamiltonYL17,FoutBSB17}, node entities have natural chemical properties. In the case of knowledge graph tasks~\cite{Nickel0TG16}, there can be a textual description of entities. When nodes have low cardinality, we can consider collaborative information. However, building models that can efficiently capture graph topology and work in a featureless setting is problematic.

Many works have attempted to address this problem. The main research direction on this topic focuses on using artificial node features. Some authors propose to initialize node features with random values~\cite{AbboudCGL21,SatoYK21}, which others use various graph statistics~\cite{Duong19,ThangHNJHA22,CuiL0Y22} and roles~\cite{ZhangC18}. You et al.~\cite{YouGYL21} proposed the use of ego-nets to embed nodes. These methods lead to an increase in the accuracy of the models, but do not solve the fundamental problem of models ignoring topological structures.

A lot of research is aimed at analyzing the expressive power of GNNs~\cite{GNNBook-ch5-li}. Xu et al.~\cite{XuHLJ19} proved that many popular message-passing models have significant limitations when it comes to learning from graph structures. They proposed a GIN model, the representational power of which is equal to the power of the 1-WL test~\cite{leman1968reduction}. Hu et al. ~\cite{HuLGZLPL20} considered the idea of transfer learning in the graph domain and proposed the GINE model, which extends GIN to handle edge attributes. Brossard et al.~\cite{Brossard20} took the expressive power of GIN-based models a step further and proposed a GINE+ model that enables the GNN to detect small cycles.


To overcome the theoretical limitations on structure learning of GNNs, researchers are turning their attention to higher-order models. Moris et al.~\cite{Morris19} and Maron et al.~\cite{Maron19a} proposed k-GNN, which associates each k-tuple of nodes with a vector representation. They showed that k-GNN is, at most, as powerful as the k-WL test. This property allows models to better understand the graph structure, but the computational complexity of such methods grows exponentially with $k$. Maron et al.~\cite{MaronBSL19} proposed PPGN, a 2-GNN based model which has a provable 3-WL expressive power. Zhu et al.~\cite{ZhuZXT21} proposed the NBFNet framework for building such models, inspired by the generalized Bellman-Ford algorithm.

Although high-order GNNs show a high possibility of learning from topology, the question of their application to heterogeneous graphs is not well studied. In our work, we propose a new second-order WalkGNN model, whose architecture allows us to efficiently account for different types of edges together with their numerical values.

\section{Algorithm}

In this section, we first present a framework for building recommendations on large graphs by analyzing ego-nets. This framework is a generalization of the idea proposed by Epasto et al.~\cite{EpastoLMSTV15}. Following them, we called it the Generalized Ego-network Friendship Score. Essentially, it represents a combination of in-ego model and out-ego aggregation.

Then, we will present a model we called WalkGNN, which we use in our framework. This model is inspired by the graph walk counting algorithm, and unlike most popular GNNs, builds representations of node pairs instead of nodes. Its key component is the novel layer WalkConv, which generates a state transition matrix for each edge and propagates the corresponding node pair states through it.

\subsection{Generalized Ego-network Friendship Score}
To illustrate our framework, consider the following scenario: each user of a social network is asked to estimate the probability of friendship between each of her or his friends based on some context. This context is the user's knowledge about his friends and the relationships between them. We can call this value the local relevance of one user to another with respect to their common friend. Then, for each pair of users, we can analyse all the local relevancies of their common friends  and infer the global relevance of one user to the other. The values obtained from this process can be used to create a list of recommended friends for each user. This idea is key to our recommendation building framework.

Instead of asking users, we will use a machine learning model that, for a pair of users $u$ and $v$, will predict their local relevance with respect to their common friend $e$. Global relevance for $u$ and $v$ is defined as an aggregation of local relevance of all their common friends. As the context of $e$, we use its ego-net. As described above, ego-nets are a very convenient structure in large-scale settings. Both scalability and informativeness of ego-nets allow building accurate predictions for hundreds of millions of users.

Our framework consists of two components. The first and the most important one we call in-ego model. This model takes an ego-net as input and produces a matrix of pairwise relevance of all nodes of this ego-net. Essentially, the in-ego model generalizes the link formation laws across graphs using their topology. The second component is the out-ego aggregation. This is just a function that reduces values for pairs of nodes, which occurs in several ego-nets. Both of these components are used in the corresponding phases of our framework.

\textbf{Phase 1: Ego-net construction.} Each ego-net is a set of triangles and pendant nodes with one common node (see~\autoref{fig:ego_net}). In order to construct ego-nets for each node of the original graph, we use a distributed version of the triangle counting algorithm augmented with a Bloom filter~\cite{Bloom70}. This algorithm can be implemented in the MapReduce paradigm. The map stage takes as input the list of node neighbors and outputs the potential triangles that passed the Bloom filter test. In the next step, we partition the triangles by the side for which we need to perform an additional check and join them with the edges of the original graph. This step is necessary to detect triangles for which the Bloom filter gave a false positive. At the final stage, we group triangles by ego and save the result. See Algorithm~\ref{alg:ego_net} for an outline of the method.

\begin{algorithm}[tb]
    \caption{Ego-nets construction}
    \label{alg:ego_net}
\begin{algorithmic}
   \STATE {\bfseries Input:} Graph $G(V, E)$
   \STATE {\bfseries Output:} $\{Ego$-$net(e)$ $|$ $\forall e \in V\}$

    \STATE $Bloom \gets BloomFitler(E)$
    \STATE $T \gets$ \O
    \FOR{$u \in V$}
        \FOR{$\forall e, v \in N(u)$}
        \IF {$(e, v) \in Bloom$}
            \STATE \textbf{Add} $(e, v, u)$ \textbf{to} $T$
        \ENDIF
        \ENDFOR
    \ENDFOR
    
    \STATE $T' \gets$ $T$ \textbf{left semi join} $E$ \textbf{on} $\lambda (x, y, z), (a, b) \Rightarrow$ $x$ = $a$ \textbf{and} $y$ = $b$
    \STATE $R \gets$ $T'$ \textbf{group by} $\lambda (x, y, z) \Rightarrow x$
    \STATE \textbf{return} $R$
\end{algorithmic}
\end{algorithm}

\textbf{Phase 2: In-ego model inference.} The ego-nets computed at the previous stage are passed to the in-ego model. This is a simple map transformation that takes an ego-net as input and outputs a matrix of pairwise relevance of nodes.

\textbf{Phase 3: Out-ego aggregation.} Each pair of nodes in the original graph will have exactly as many scores as they have common neighbors. To obtain the final relevance value between two nodes, the out-ego aggregation function is used. In this work, we use simple binary operations such as summation or maximum. The resulting values can be used to construct friend suggestions.


\textbf{Note.} Some classic link prediction heuristics fit our framework. In case of Adamic-Adar, an in-ego model is a value $\dfrac{1}{\log(n)}$, where $n$ is the size of the ego-net, and out-ego aggregation is summation. In case of friendship score~\cite{EpastoLMSTV15}, an in-ego model's value is $1$ when both nodes are in the same cluster, and $0$ otherwise.

All phases of our framework can be implemented in the MapReduce paradigm. This makes it possible for us to scale it to graphs of any size. A key component of our framework is the in-ego model. Its accuracy directly affects the quality of the entire system. In the next section, we will describe our GNN, which we developed specifically for our task by taking into account its specifics.

\subsection{WalkGNN}
Our framework can easily scale to graphs of any size, but this property also imposes certain limitations, since in-ego models must consider all ego-nets independently of each other. In other words, natural node identifiers cannot be used within the model, and therefore, collaboration information is unavailable. The edges in our task change dynamically and can be of different types such as friendships, messages, likes and others. Finally, there are no natural features of users that would provide sufficient information about their social connections' dynamics. Only general features such as sex, age or country are available. These kinds of features provide only auxiliary information and cannot be a reliable basis. Thus, the problem can be described as graph-level link prediction on heterogeneous temporal graphs without node attributes.

The lack of node attributes in the graph-level setting limits model design. Therefore, we have two options available. The first is to build the model in the node representation paradigm and craft synthetic node features. Such methods have been widely studied by the community and have shown a strong dependence on hand-crafted features. In contrast, the second option of models based on node pair representations is much less studied, but existing works show promising results~\cite{ZhuZXT21}. Morris et al.~\cite{Morris19} and Maron et al.~\cite{MaronBSL19} proved that the k-GNN model is at most as powerful as the k-WL test. Since the ability of the model to efficiently learn the graph structure is critical to our problem, we built our model based on the node pair representations paradigm.

Our WalkGNN model is inspired by the walks counting algorithm. This algorithm is a simple example of a dynamic programming problem. To calculate the number of walks of length $k$ between all pairs of nodes in a graph, we must first solve the same problem for length $k-1$ and multiply the result by the adjacency matrix: $W_k = W_{k-1} \times A$, where $W_k$ is a $k$-length walks count matrix and $A$ is the adjacency matrix. The initial state is an identity matrix.

Instead of a scalar value of the walks count, we will store a vector of size d, which is an embedding that describes the relationship between a pair of nodes. To predict the probability of forming links, we can train a supervised classifier based on them. To construct such embeddings, we need to define an initial state and a transition function. We can use a diagonal matrix with constant values or features of nodes as the initial state. The transition function will be defined in the next subsection.

\textbf{WalkConv.} Suppose we have representations of node pair relations at the $k$-th iteration. That is, the state matrix $W_k$ of dimensions $[n \times n \times d]$ is computed, and we need to compute $W_{k+1}$. A key principle of our architecture is that we treat each edge as an information filter through which we propagate the state. Remember that each edge is parameterized by different attributes, such as the age of the friendship, the number of messages, likes or profile visits, and so on. In general, for each edge, we have a vector of its attributes $\overline{e}$. On top of it, we can build a Multilayer perceptron (MLP), which would transform $\overline{e}$ into a $d^2$-dimensional vector. It can later be reshaped to a $[d \times d]$ dimensional matrix. The obtained matrix is the linear filter used for propagating the state. The transition formula is as follows:
\begin{equation}
    W_{k+1}^{u, v} := \dfrac{1}{d} \sum\limits_{(t, v, \overline{e}) \in E} W_{k} ^ {u, t} \times EdgeMLP_k(\overline{e})
\end{equation}
where $d$ is the dimension of the state, $E$ is a set of edges and $EdgeMLP_k$ is an MLP that takes the attributes of an edge as input and returns a matrix of dimension $[d \times d]$. This expression can also be represented in Einstein notation, as presented in Algorithm~\ref{alg:walkconv}. The complexity of WalkConv is $\mathcal{O}(n^3 \times d^2)$, where $n$ is the number of nodes and $d$ is the number of hidden units. This is the most computationally intensive part of our model, but it can be implemented efficiently as a dense multiplication of two matrices of dimensions $[n \times n \times d]$ and $[n \times n \times d \times d]$.

\begin{algorithm}[tb]
    \caption{WalkConv}
    \label{alg:walkconv}
\begin{algorithmic}
    \STATE {\bfseries Input:} current state matrix $S_k$, edge list $E$
    \STATE {\bfseries Output:} new state matrix $S_{k+1}$
    \STATE $T_k^{u,v} \gets EdgeMLP_k(\overline{e}), (u, v, \overline{e}) \in E$
    \STATE $W_{k+1}^{u,v,t} \gets \dfrac{1}{d} S_k^{u,q,c} T_k^{q,v,c,t}$
    \STATE \textbf{return} $S_k + MLP_k(W_{k+1})$
\end{algorithmic}
\end{algorithm}

The WalkGNN model itself is a stack of WalkConv layers. To improve learning stability, we also use a residual connection after each layer (see Algorithm~\ref{alg:walkconv} and~\ref{alg:walkgnn} for
pseudocode).

\begin{algorithm}
    \caption{WalkGNN}
    \label{alg:walkgnn}
\begin{algorithmic}
    \STATE {\bfseries Input:} edge list $E$, number of layers $l$
    \STATE {\bfseries Output:} pairwise matrix relevance

\STATE $S_0^{u,v} \gets     \begin{cases}
            \overline{1}, &         \text{if } u=v,\\
            \overline{0}, &         \text{if } u\neq v.
    \end{cases}$

\STATE  \FOR{$k \gets 1$ to $l$}
        \STATE {$S_k \gets WalkConv(S_{k-1}, E)$}
        \ENDFOR

\STATE \textbf{return} $OutMLP(S_l)$

\end{algorithmic}
\end{algorithm}

\section{Experiment}
Firstly, we compare the accuracy of the WalkGNN\footnote{Code and dataset are provided at: \gitlink{}.} model with classical heuristics as well as some of the most diffused models in the literature that are applicable in our setting. Then, we describe the results of an online test based on \VK{}'s friends recommendation system. Finally, we show the results of an ablation study that validates the importance of each model component in maximizing accuracy.

\subsection{Offline Experiment}

\textbf{Dataset.}
We evaluate our model on two different datasets. The first dataset, \egovk{}, is a random sample of ego-nets from users of the \VK{} social network for a certain date. The friendships formed on the next day are used as the ground truth edges, ego-nets with no new friendships during this period are ignored. Ground truth edges are undirected, i.e., any direction of the edge is considered as the right prediction. The base edges inside ego-nets are directed and can be of four types. Each type of edge have numerical characteristic describing the age of the friendship or three other kinds of activity. If there is no friendship between a pair of nodes, the age value is $-1$. It is guaranteed that there are no base edges of any type between pairs of nodes from the ground truth set in the original ego-net. The training, validation and test set are divided randomly. Natural node attributes are not used in this task. As attributes of nodes, we use attributes of edges with ego node in forward and backward directions. The edges with ego node at one end are omitted. In each ego-net, the ego node has id $0$ and all other nodes are numbered with ordinal numbers, so all ego-nets are completely independent. All ego-nets are limited by $300$ nodes by the activity of interactions with ego.

To evaluate the applicability of our model to graphs of other domains, we also measure its accuracy on the Yeast~\cite{YanCHY08, TUDataset} dataset. This dataset is a set of small molecules, where each molecule is described as a graph with labeled nodes and edges. The nodes are described by one of 74 labels, and the edges are of three types. Originally, this dataset represents a graph-level classification task. To evaluate the quality of the link prediction task, we adapted the dataset accordingly. We removed one edge from each graph and assigned it as a target for the model. Note that graph labels are not used in our task. The statistics of both datasets are provided in~\autoref{tab:stats}.

\begin{table}[t]
\caption{\egovk{} dataset stats.}
\label{tab:stats}
\begin{center}
\begin{small}
\begin{sc}
\begin{tabular}{ccccc}
\toprule
    Dataset & Graphs & Avg. nodes & Avg. edges & Avg. labels \\ 
\midrule
 \egovk{} & 61808 & 210.4 & 2916.6 & 1.5 \\ 
 Yeast & 79601 & 21.5 & 22.8 & 1.0 \\ 
 \bottomrule
\end{tabular}
\end{sc}
\end{small}
\end{center}
\end{table}

\textbf{Evaluation.} We use the ndcg@5 measure to compare the link-prediction accuracy of the models. In each ego-net, the $5$ pairs of nodes with best scores other than those already present are used to score the model. The hyperparameters of the models were selected to maximize accuracy on the validation dataset. Accuracies of the models on the test sample together with 95\% confidence intervals
are presented in~\autoref{tab:results}.

\textbf{Implementation Details.}
In our WalkGNN model, we use $6$ WalkConv layers with $8$ hidden units. We also extend the attributes of edges by adding the features of the source and destination nodes of the corresponding edge. In the case of directed graphs, at the end of WalkConv, we concatenate the new state with its transpose to consider the reverse direction of edges. The time feature from \egovk{} dataset is transformed by formula $28/(t+1)$ when $t \geq 0$ and $0$ otherwise. Attributes of edges between ego and the node in both directions are used as features of this node at the initialization step. In the rest of the model, the edges with ego are ignored. All MLPs have $4$ layers, ReLU activation and $32$ hidden units. As a loss function, we use pairwise loss~\cite{BurgesSRLDHH05}.

\textbf{Baselines.}
We compare WalkGNN against classical heuristics and the most popular GNNs that are known to be able to learn from the graph structure. As a heuristics baseline, we use Adamic-Adar and its weighted version. We use the state-of-the-art GIN model and its GINE modification as the baselines, which are known in the literature as the most expressive MP-GNN. Xu et al.~\cite{XuHLJ19} and Errica et al.~\cite{ErricaPBM20} argue that using node degree as a feature with GIN improves model accuracy. Sato et al.~\cite{SatoYK21} introduce rGIN model which assign additional attributes of each node sampled from an almost uniform discrete probability distribution. Unfortunately, this modification does not improve the accuracy of the model on our task. Other first-order models such as GCN, GraphSage and GAT failed on the task. We use PPGN~\cite{MaronBSL19} as a high-order GNN baseline, which has strong theoretical expressive power, high accuracy in link prediction problems, a simple and clear architecture, and the ability to incorporate edge attributes. Higher-order GNNs are computationally expensive and are not suitable for our problem. In order to correctly compare expressive power, we consider a setting without edge attributes as well.

\textbf{Results.}
The results of model evaluation are presented in~\autoref{tab:results}. WalkGNN outperforms the next-best solution by \textbf{20\%} and \textbf{3\%} in the setting with and without edge attributes on the \egovk{} dataset. It is worth noting that Adamic-Adar is still a strong baseline in social networks domain and outperforms first-order GNN in terms of accuracy, which is explained by the lack of dependence on the node features. The 156\% gap between WalkGNN versions with and without edge attributes shows the model's high degree of attention to graph structure and edge properties. On featureless tasks, the superiority of WalkGNN over Adamic-Adar, which is widely used in social networks, demonstrates the model's ability to efficiently learn the natural laws of edge formation.

The evaluation results on the Yeast dataset also show the superiority of our model over baselines. The low result of the Adamic-Adar heuristic shows that the laws of link formation in the social domain are inapplicable to chemical bond analysis. Despite this, WalkGNN performs well and significantly outperforms other baseline models in terms of accuracy.

\begin{table}[t]
\caption{Test set ndcg@5.}
\label{tab:results}
\begin{center}
\begin{small}
\begin{sc}
\footnotesize{
\begin{tabular}{lccc}
\toprule
    Algorithm & \egovk{}-no-attr & \egovk{} & Yeast \\ 
\midrule
 Adamic-Adar &     0.028 $\pm$ 3$\cdot$$10^{-3}$ &          0.055 $\pm$ 4$\cdot$$10^{-3}$  & 0.003 $\pm$ 7$\cdot$$10^{-4}$ \\ 
 GIN &             0.015 $\pm$ 2$\cdot$$10^{-3}$ &          0.037 $\pm$ 3$\cdot$$10^{-3}$  & 0.405 $\pm$ 6$\cdot$$10^{-4}$ \\
 GIN-deg &         0.015 $\pm$ 2$\cdot$$10^{-3}$ &          0.037 $\pm$ 3$\cdot$$10^{-3}$  & 0.407 $\pm$ 7$\cdot$$10^{-4}$ \\
 PPGN &            0.034 $\pm$ 3$\cdot$$10^{-3}$ &          0.075 $\pm$ 4$\cdot$$10^{-3}$  & 0.642 $\pm$ 6$\cdot$$10^{-4}$ \\
 WalkGNN & \textbf{0.035} $\pm$ 3$\cdot$$10^{-3}$ & \textbf{0.090} $\pm$ 4$\cdot$$10^{-3}$ & \textbf{0.720} $\pm$ 6$\cdot$$10^{-4}$ \\
\bottomrule
\end{tabular}}
\end{sc}
\end{small}
\end{center}
\end{table}

\subsection{Online Experiments.}
\begin{figure}[h]
  \centering
  \includegraphics[width=0.75\linewidth]{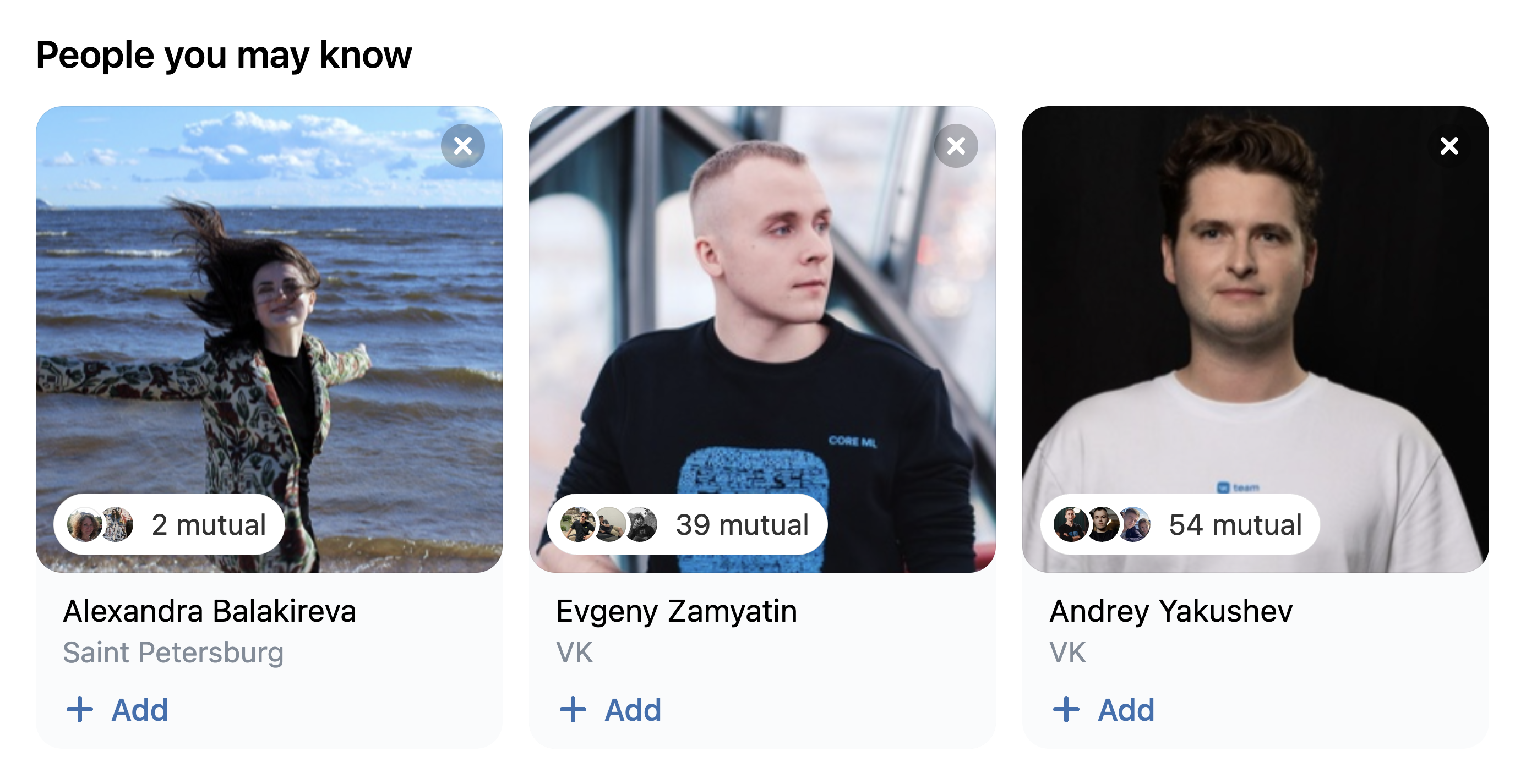}
  \caption{PYMK at \VK{}.}
 .\label{fig:pymk}
\end{figure}

We test Generalized Ego-network Friendship Score with the WalkGNN model in a People You May Know (PYMK) recommendation block of the \VK{} social network (see~\autoref{fig:pymk}). The block is shown to the users of the news feed with more than \mau{} monthly users. To compare our method with the production solution, we divide this audience into two parts for the AB test. The algorithm on the control group is a modified version\footnote{See \mediumlink{} for details.} of the friendship score~\cite{EpastoLMSTV15}, which outperformed our previous solutions such as Adamic-Adar and Node2Vec\footnote{\nodetoveclink{}} with a large margin. Both methods work offline and build and export suggestions for all \VK{} users every day. Pipeline consists of three parts: ego-nets materialization, per ego-net computation and results aggregation. All stages together take up about 3k cores and 3Tb RAM of a hadoop~\cite{ShvachkoKRC10} cluster and take about 10 hours to complete.
The results of the online test showed a \textbf{12\%} increase in friend requests' CTR.

\subsection{Ablation Study.}
To prove the importance of components of the WalkGNN model, we perform a series of ablation experiments on the \egovk{} dataset. We take the model described in Section 4.1 as the base configuration and make one change in each experiment:
\begin{itemize}
    \item Use $2$ WalkConv blocks instead of $6$
    \item Use $4$ WalkConv blocks instead of $6$
    \item Use $8$ WalkConv blocks instead of $6$
    \item Ignore node features
    \item Ignore edge features
\end{itemize}

\textbf{Results.} The results of the experiments are presented in \autoref{tab:ab_results}. The accuracy of the model grows with the number of convolution blocks, and saturates around 6 on our task. 
The edge attributes have a significant impact on the accuracy of the model, as the performance degradation when ignoring them is more than 50\%. In contrast, the results of ignoring node features show the insignificance of their influence on model performance due to their low quality or inefficient use. We leave the study of how to effectively utilize node features for future works.

\begin{table}[ht]
\caption{Ablation study results.}
\label{tab:ab_results}
\begin{center}
\begin{small}
\begin{sc}
\begin{tabular}{lc}
\toprule
Algorithm & ndcg@5 \\
\midrule
     WalkGNN-base & 0.090 $\pm$ 4$\cdot$$10^{-3}$ \\
     WalkGNN-2b & 0.074 $\pm$ 4$\cdot$$10^{-3}$ \\
     WalkGNN-4b & 0.085 $\pm$ 4$\cdot$$10^{-3}$ \\
     WalkGNN-8b & 0.088 $\pm$ 4$\cdot$$10^{-3}$ \\
     WalkGNN-no-e-attr & 0.051 $\pm$ 3$\cdot$$10^{-3}$ \\
     WalkGNN-no-n-attr & 0.086 $\pm$ 4$\cdot$$10^{-3}$ \\
\bottomrule
\end{tabular}
\end{sc}
\end{small}
\end{center}
\end{table}

\section{Conclusion}
In this paper, we presented a framework for building friend suggestions on large-scale graphs. Our solution is capable of leveraging the accuracy of complex models while maintaining scalability. As part of our framework, we addressed the graph-level problem of predicting friendships within ego-nets. We developed the WalkGNN model that satisfies all the requirements of our framework and is capable of outperforming state-of-the-art solutions. To measure the quality metrics relevant to our task, we introduced the \egovk{} dataset, which represents a supervised graph-level link-prediction task. We believe that our dataset could be useful for further research addressing the expressive power of graph models.

An interesting direction for future work is to investigate more complex local relevance aggregation methods that can be extended from scalar values to multidimensional vectors. Another option is to develop a way to efficiently incorporate node attributes into the WalkGNN architecture.

\bibliographystyle{unsrt}  
\bibliography{main}
\end{document}